\documentclass{appolb}
\usepackage{epsfig}
\newcommand{\beq}{\begin{equation}}
\newcommand{\eeq}{\end{equation}}
\newcommand{\bea}{\begin{eqnarray}}
\newcommand{\eea}{\end{eqnarray}}
\def\la{\langle}
\def\ra{\rangle}
\def\al{\alpha}

\def\gam{\gamma}

\def\half{\frac{1}{2}}
\def\la{\langle}
\def\ra{\rangle}
\def\al{\alpha}

\def\gam{\gamma}

\def\half{\frac{1}{2}}
\def\nubar{\overline{\nu}}

\def\GEem{G_E}
\def\GMem{G_M}

\def\GENC{G_E^{NC}}
\def\GMNC{G_M^{NC}}
\def\GANC{G_A^{NC}}
\def\GACC{G_A}

\def\GMCC{G_M^{CC}}
\def\GECC{G_E^{CC}}

\def\GAs{G_A^s}

\def\GMs{G_M^s}
\def\GEs{G_E^s}
\def\Acal{\mathcal{A}}
\def\Ncal{\mathcal{N}}

\begin{document}
\eqsec  
\title{Neutrino scattering as a probe for the strange content of the nucleon
\thanks{e-mail address:~{\tt alberico@to.infn.it}}
}
\author{W.M. Alberico
\address{Dip. di Fisica Teorica, University of Torino\\
 and INFN, Sezione di Torino, via P.Giuria 1, Torino, Italy}}

\maketitle
\begin{abstract}
We consider different methods and observables which can be obtained by the 
measurement of neutrino scattering off nucleons and nuclei with the purpose of
finding evidence for the strange form factors of the nucleon, which enter into 
structure of the nucleonic weak neutral current.
\end{abstract}
  
\section{Introduction}

The contribution of the $s\bar{s}$ sea to the nucleon structure has been widely
discussed in the recent past, especially in connection with the so-called 
``problem of the proton spin''. It is related to the one nucleon matrix element 
of the axial quark current:
\bea
\la p,s|\bar{q}\gam^\al\gam^5 q|p,s\ra = 2Ms^\al {g^q_A},
\label{matrixel}
\eea
$q,\bar{q}$ being the quark fields and $|p,s\ra$ the proton (momentum, spin) 
state vector.

Based on several assumptions, among which the use of the naive parton model (thus
neglecting important QCD corrections) and SU(3) flavor symmetry, the constants 
$g^u_A, g^d_A, g^s_A$ can be determined from:
\begin{enumerate}
\item
{QCD sum rule} (of the polarized structure function)
\[\Gamma_1^p=\int_0^1dx g_1^p(x)
=\half \left(\frac{4}{9}\Delta u +\frac{1}{9}\Delta d +
\frac{1}{9}\Delta s\right)\]
\item
{the relation $g_A=g_A^u-g_A^d$}\\
with $g_A=1.2573\pm 0.0028$ obtained from neutron decay
\item
{the relation $3F-D= g_A^u+g_A^d-2g^s_A$}\\
where the constants $F,D$ are measured from semileptonic decays of hyperons.
\end{enumerate}
At face with the above mentioned theoretical uncertainties, it is highly desirable
to find an alternative source of information  for an independent 
determination of $g^s_A$ ~\cite{ABM-report}.

In this perspective a very important tool is the measurement of neutral current 
(NC) neutrino cross sections
\begin{equation}
\nu_{\mu} (\nubar_{\mu}) + N \longrightarrow
\nu_{\mu} (\nubar_{\mu}) + N\, ,
\label{nuelas}
\end{equation}
 together with the charge current (CC) processes:
{\begin{equation}
\begin{array}{l}
\nu_{\mu}+n\longrightarrow \mu^-+p\,,\\
\nubar_{\mu}+p\longrightarrow \mu^+ +n\,.
\end{array}
\label{nuquasiel}
\end{equation}}
The neutral current involved in the above processes is:
\beq
J_\al^Z\equiv J_\al^{NC}=V_\al^3+A_\al^3-2\sin^2\theta_W J^{em}_\al-
\frac{1}{2}V^s_\al-\frac{1}{2}A^s_\al,
\label{NCformal}
\eeq
with $V_\al^3=\bar{U}\gam_\al U- \bar {D}\gam_\al D$, 
$A_\al^3=\bar{U}\gam_\al\gam_5 U- \bar {D}\gam_\al\gam_5 D$, 
$V_\al^s=\bar{S}\gam_\al S$ and $A_\al^s=\bar{S}\gam_\al\gam_5 S$.
The charge current reads, instead:

\beq
J_\al^W=V_{ud}\bar{U}\gam_\al(1+\gam_5)D.
\label{CCformal}
\eeq

The one-nucleon matrix elements of the above currents are usually expressed in terms 
of phenomenological form factors, which contain, in the NC sector, three isoscalar 
strange terms, $\GEs$, $\GMs$ and $G_A^s$. It is our purpose to show how these form
factors can be determined from $\nu$ ($\bar\nu$) scattering processes.

\section{neutrino-nucleon (-nucleus) cross sections}
\subsection{Elastic NC $\nu$--nucleon scattering}
The differential cross section for the elastic NC $\nu$-nucleon scattering 
reads \cite{ABM-report}
\bea
\!\!\!\!\!\!\!\!\!\!
&&\left(\frac{d\sigma}{dQ^2}\right)^{NC}_{\nu(\nubar)} = 
\frac{G_F^2}{2\pi}\left[\half y^2(\GMNC)^2 
+\left(1-y-\frac{M}{2E}y\right)
\frac{{(\GENC)^2+\frac{E}{2M}y(\GMNC)^2}}
{{1+\frac{E}{2M}y}}\right.
\nonumber\\
&&\,\left. +\left(\half y^2+1-y+\frac{M}{2E}y\right)(\GANC)^2
\pm 2y\left(1-\half y\right)\GMNC\GANC\right]\,,\!\!\!\!\!\!\!
\label{elasnc}
\eea
the + (-) sign referring to neutrinos (anti-neutrinos) respectively. In the above
$y=\frac{p\cdot q}{p\cdot k} =\frac{Q^2}{2p\cdot k}$, $E$ is the energy of the 
 neutrino (antineutrino) in the laboratory system and the NC Sachs form factors 
can be expressed as:
\bea
&&{G_E}^{NC;p(n)}(Q^2) =
\label{GEnc}\\
&&\quad = \pm\half\left\{\GEem^p(Q^2)-\GEem^n(Q^2)\right\}-
2\sin^2\theta_W\GEem^{p(n)}(Q^2) -\half{\GEs(Q^2)}
\nonumber\\
&&{G_M}^{NC;p(n)}(Q^2) = 
\label{GMnc}\\
&&\qquad = \pm\half\left\{\GMem^p(Q^2)-\GMem^n(Q^2)\right\}-
2\sin^2\theta_W\GMem^{p(n)}(Q^2) -\half{\GMs(Q^2)}
\nonumber\\
&&G_A^{NC;p(n)}(Q^2)=\pm\half \GACC(Q^2) -\half {G_A^s(Q^2)}.
\label{GAnc}
\eea
They clearly embody the isoscalar contribution of the electric ($\GEs$), 
magnetic ($\GMs$) and axial ($G_A^s$) strange form factors.

\subsection{Quasi-Elastic CC $\nu$--nucleon scattering}

The differential cross section for the quasi-elastic CC $\nu$-nucleon scattering 
reads \cite{ABM-report}
\bea
\!\!\!\!\!\!\!\!\!\!&&\left(\frac{d\sigma}{dQ^2}\right)^{CC}_{\nu(\nubar)} =
\frac{G_F^2}{2\pi}\left[\half y^2(\GMCC)^2 +\left(1-y-\frac{M}{2E}y\right)
\frac{{(\GECC)^2+\frac{E}{2M}y(\GMCC)^2}}
{{1+\frac{E}{2M}y}}\right.+
\nonumber\\
&&\,\left.
+\left(\half y^2+1-y+\frac{M}{2E}y\right)(\GACC)^2
\pm 2y\left(1-\half y\right)\GMCC\GACC\right]\,.\!\!\!\!\!\!\!
\label{quasiel-cc}
\eea
In the above the CC form factors appear.

\subsection{Quasi-Elastic  $\nu$--nucleus scattering}

Neutrino scattering is realized both on free and bound nucleons, often within the 
same experiment/detector: it is thus important to consider also $\nu$--nucleus 
scattering processes and to accurately evaluate the effects of nuclear structure and 
dynamics.

The processes on a nucleus are, as before, of two types:
\bea
\nu_\mu({\overline\nu_\mu}) + A &&\longrightarrow
\nu_\mu({\overline\nu_\mu}) +N + (A-1)\qquad\quad\mathrm{NC\,\, process}
\label{nucleusNC}\\
\nu_\mu({\overline\nu_\mu}) + A &&\longrightarrow
\mu^-({\mu^+}) +p(n) + (A-1)\, \qquad\mathrm{CC\,\, process}
\label{nucleusCC}
\eea
The approach one usually employs is the Impulse Approximation (IA), 
in which the neutrino 
interacts with a single nucleon in the nucleus; the latter can be described within
the simplest available model, namely the (relativistic) Fermi Gas (RFG) or within a
more refined relativistic shell model (RSM) or taking into account initial state 
correlations (RPA and the like); an initial binding energy of the struck nucleon is 
customarily taken into account, also in RFG, though its effect is negligible for 
neutrino energies in the GeV range.

After weakly interacting with the neutrino, the nucleon ejected in the processes 
(\ref{nucleusNC}) and (\ref{nucleusCC}) can be treated either as a free one (PWIA) or 
as interacting with the residual nucleus (DWIA). Different approaches are available 
in order to deal with the distortion of the final nucleon [See more on this subject 
in Ref.\cite{Maieron,ABBCGM}].

In spite of its simplicity, the RFG model turns out to be useful, both in evaluating 
ratios of cross sections, where a large part of the nuclear effects fade away, and to 
get a feeling of how the various nucleonic form factors (including the strange ones) 
enter into the game. Indeed the RFG cross sections can be analytically evaluated and
read:
\begin{eqnarray}
\left(
\frac{d^2\sigma}{d E_N d\Omega_N}\right)_{\nu(\overline\nu)}
&&=
\frac{G_F^2}{(2\pi)^2} \frac{3 \Ncal}{4\pi p_F^3}\frac{|{\vec{p}}_N|}{k_0 }
\int \frac{d^3 k'}{k_0'}\frac{d^3 p}{p_0}\delta\left(k_0-k_0'+p_0-E_N\right)
\nonumber\\
&&\times \delta^{(3)}\left({\vec{k}}-{\vec{k}}'+{\vec{p}}-{\vec{p}}_N \right)
\theta(p_F-|{\vec{p}}\,|)\theta(|{\vec{p}}_N|-p_F)
\nonumber\\
&&\times \left\{ V_M (\GMNC)^2 + V_{EM} 
\frac{(\GENC)^2 + \tau (\GMNC)^2}
{1 + \tau} +\right.
\nonumber\\
&&+ \left. V_A (\GANC)^2 \pm V_{AM} \GANC\GMNC \right\},
\label{RFGcs}
\end{eqnarray}
where $(\vec{p_N}, E_N)$ is the four momentum of the ejected nucleon, $k$ ($k'$) 
the momenta of the incoming (outgoing) neutrino and
\begin{equation}
\begin{array}{l}
V_M = 2 M^2 \tau \left(k \cdot k^{\prime}\right) \\
V_{EM} = 2 \left(k \cdot p\right) \left(k^{\prime} \cdot p\right)
  - M^2 \left(k\cdot k^{\prime}\right) \\
V_A = M^2 \left(k\cdot k^{\prime}\right) +
 2 M^2 \tau \left(k \cdot k^{\prime}\right)  + 2 \left(k \cdot p \right)
 \left(k^{\prime} \cdot p\right)  \\
V_{AM} = 2 \left(k\cdot k^{\prime}\right) 
\left(k \cdot p + k^{\prime}\cdot p \right)\, ,
\end{array}
\nonumber
\end{equation}
the remaining symbols being self-explanatory.

The single differential cross sections then follow:
\begin{equation}
\left(\frac{d\sigma}{d T_N}
\right)_{\nu(\overline\nu) N}\equiv
\left(\frac{d\sigma}{d E_N}
\right)_{\nu(\overline\nu) N} =
\int\, d\Omega_N 
\left(\frac{d^2\sigma}{d E_N d\Omega_N}\right)_{\nu(\overline\nu)N}\;,
\label{RFGcs2}
\end{equation}
$T_N$ being the outgoing nucleon kinetic energy.

\section{Interesting observables}

As it has been suggested several times \cite{ABM-report}, the extraction of information
on the strange form factors of the nucleon from $\nu$-nucleon or $\nu$-nucleus 
scattering cross section is better founded on the measurement of {\bf ratios} of 
cross sections. This limits some of the experimental uncertainties as well as (in 
the case of $\nu$-nucleus scattering) of the model dependence of the calculated 
cross sections.

Hence the following ratios have been proposed:
\begin{itemize}
\item
NC over CC ratio (considered at Fermilab):
\begin{equation}
R_{NC/CC}(Q^2)= 
{\displaystyle{\left({d\sigma}/{dQ^2}\right)^{NC}_{\nu}}}/
{\displaystyle{\left({d\sigma}/{dQ^2}\right)^{CC}_{\nu}}}
\label{NC-CC}
\end{equation}

\item
Proton to neutron ratio (in quasi-elastic processes with emission
of one nucleon)~\cite{ABBCGM98}:
\begin{equation}
R_{p/n}^{\nu}(Q^2)= 
{\displaystyle{\left(\frac{d\sigma}{dQ^2}\right)^{NC}_{(\nu,p)}}}/
{\displaystyle{\left(\frac{d\sigma}{dQ^2}\right)^{NC}_{(\nu,n)}}}
\label{Rpn}
\end{equation}

\item
Neutrino-antineutrino Asymmetry~\cite{ABGM}:
\begin{equation}
\Acal(Q^2)= \frac{\displaystyle{\left(\frac{d\sigma}{dQ^2}\right)^{NC}_{\nu} -
\left(\frac{d\sigma}{dQ^2}\right)^{NC}_{\nubar}}}
{\displaystyle{\left(\frac{d\sigma}{dQ^2}\right)^{CC}_{\nu} -
\left(\frac{d\sigma}{dQ^2}\right)^{CC}_{\nubar}}}\,
\label{nuasymm}
\end{equation}

\end{itemize}

Since mono-energetic neutrinos are not available, in the above combinations it 
is customary to employ flux averaged neutrino cross sections:
\begin{equation}
\langle\frac{ d\sigma}{d Q^2}
\rangle^{NC}_{\nu (\nubar)} =
\frac{ \displaystyle{\int d E_{\nu (\overline{\nu})} 
\left(d\sigma/dQ^2\right)^{NC}_{\nu (\nubar)} 
\Phi_{\nu(\overline{\nu})} 
\left( E_{\nu (\overline{\nu})} \right)} }
{ \displaystyle{\int d E_{\nu(\overline{\nu})}
\Phi_{\nu(\overline{\nu})} 
\left( E_{\nu (\overline{\nu})} \right)} }\;,
\label{fluxave}
\end{equation}
the flux $\Phi_{\nu(\overline{\nu})}(E)$ being provided by the various experiments.

\subsection{The $\nu-\bar\nu$ asymmetry}

The neutrino-antineutrino asymmetry in $\nu$($\bar\nu$)-nucleon
elastic scattering explicitly reads~\cite{ABGM}:
\bea
&&\Acal_{p(n)}=\frac{1}{4}\left(\pm 1-\frac{\GAs}{\GACC}\right)
\left(\pm 1-2\sin^2\theta_W\frac{{\GMem}^{p(n)}}{\GMem^3} 
-\half\frac{\GMs}{\GMem^3}\right)\,.
\label{nuasymm2}
\eea

Thus, in the asymmetry $\Acal$ the strange axial and vector form 
factors enter in the form of  ratios, ${\GAs}/{\GACC}$ and
${\GMs}/{\GMem^3}$. Taking into account only terms which linearly depend on the strange
form factors one gets: 
\beq
\Acal_{p(n)}=\Acal_{p(n)}^0 \mp\frac{1}{8}\frac{{\GMs}}{\GMem^3} 
\mp \frac{{\GAs}}{\GACC}{\Acal_{p(n)}^0}
\label{nuasymm3}
\eeq
with $\Acal_{p(n)}^0= \frac{1}{4}\left(1\mp 
2\sin^2\theta_W\frac{{\GMem}^{p(n)}}{\GMem^3} \right)$.
%

\begin{figure}[ht]
\begin{center}
\mbox{\epsfig{file=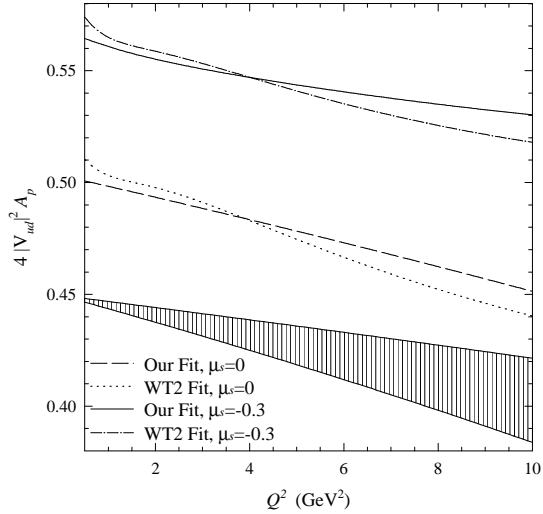,width=0.55\textwidth}}
\end{center}
\caption{Plot of $4|V_{ud}|^2{\Acal}_p$ as a function of $Q^2$. In the upper lines 
$g_A^s=-0.15$.}
\end{figure}

The quantity (\ref{nuasymm3}) is plotted in Fig.1, where the shaded area 
(corresponding to the 
uncertainty in the magnetic form factors) is the result obtained without strange 
form factors. Should the latter be different from zero, their effect would be 
clearly visible.

\section{The BNL - 734 experiment}

In 1987 the 734 experiment at BNL measured the following ratios~\cite{Ahrens87}:
\bea
R_\nu &=& 
\frac{\langle \sigma \rangle_{(\nu p\rightarrow \nu p)}}
{\langle \sigma \rangle_{(\nu n\rightarrow
\mu^- p)}} = 0.153 \pm 0.007 \pm 0.017
\label{Rnu} \\
R_{\overline{\nu}} &=& 
\frac{\langle \sigma \rangle_{(\overline{\nu} p\rightarrow 
\overline{\nu} p)}}
{\langle \sigma \rangle_{(\overline{\nu} p\rightarrow
\mu^+ n)}} = 0.218 \pm 0.012 \pm 0.023
\label{Rnubar} \\
R &=& 
\frac{\langle \sigma \rangle_{(\overline{\nu} p\rightarrow 
\overline{\nu} p)}}
{\langle \sigma \rangle_{(\nu p\rightarrow
\nu p)}} = 0.302 \pm 0.019 \pm 0.037\ ,
\label{RR}
\eea
where $\langle \sigma \rangle_{\nu(\bar\nu)}$ is a total cross section
integrated  over the incident neutrino (antineutrino) energy and weighted 
by the $\nu(\bar\nu)$ flux. The first error is statistical and the second is 
the systematic one.

In terms of these ratios, the {``integrated'' asymmetry} reads:
\beq
\langle {\cal A}_p \rangle =
\frac{R_\nu(1-R)}{1-RR_\nu/R_{\overline{\nu}}}
\label{nuasymexp}
\eeq
and from the  experimental data we found \cite{ABBCGM99}
\beq
\langle {\cal A}_p \rangle = 0.136 \pm 0.008 (\mathrm{stat}) 
\pm 0.019 (\mathrm{syst}),
\label{nuasymexp2}
\eeq
which is the only existing measurement of the neutrino asymmetry.
The RFG ratios (\ref{Rnu})--(\ref{RR}) are plotted in Fig.2, together with the 
asymmetry, as functions of the magnetic strangeness and for different choices of 
parameters entering into the axial ($g_A^s$) and electric ($\rho_s$) strange form 
factors [for a parameterization of the latter see Ref.\cite{ABM-report}]. The shaded
areas correspond to the experimental band.

\begin{figure}[ht]
\begin{center}
\mbox{\epsfig{file=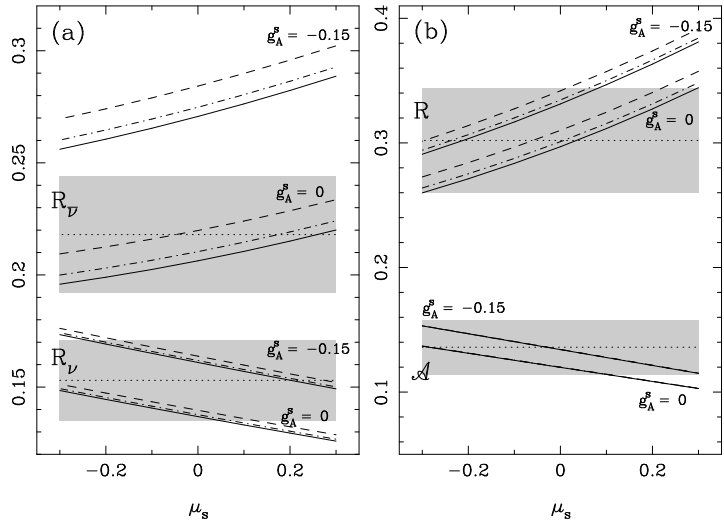,angle=-90,width=0.85\textwidth}}
\end{center}
\caption{ $R_{\nu}$ and $R_{\bar\nu}$, $R$ and $\langle {\cal A}_p\rangle$, 
 for $g_A^s=0$ and $g_A^s=-0.15$. Three choices of $\rho_s$ are 
shown: $\rho_s=0$ (solid),$\rho_s=-2$ (dot--dashed) and $\rho_s=+2$ (dashed).
}
\end{figure}

The figure clearly shows the sensitivity of the various ratios to the strange form 
factors: notice that the quantities (\ref{Rnu}) and (\ref{Rnubar}) are ratios of NC 
over CC cross sections. Unfortunately the error band of the BNL-734 experiment are 
too large to draw definitive conclusions, but they certainly do not exclude relevant
contribution of the strange sea to the nucleon structure.

\section{Conclusions and future perspectives}

The experiments of $\nu$-proton NC and CC scattering are {highly
interesting} for the determination of $\Delta s\equiv g_A^s$. Among the various
proposed observables, the ration of NC and CC elastic $\nu p$ scattering cross 
sections is being considered in several proposal at Fermilab (Minerva, FINESSE).

This quantity will be sensitive to $g_A^s$, but possibly not much affected by the 
cutoff mass of the axial form factors (assuming, for simplicity, the same dipole 
form in the strange form factors as in the ordinary ones). Also the e.m. form factors 
do not sensibly affect the NC/CC ratio. 

One of the major uncertainties in the extraction of $g_A^s$ from $\nu$-N scattering 
stems from the unavoidable interference between the axial and the magnetic ($\mu_s$)
strange form factors. Neutrino scattering alone can only determine a linear combination
of both; hence it is highly desirable to obtain a complementary information on $\mu_s$ 
from a different source. This is the case of parity violating (PV) scattering of 
polarized electrons from protons~\cite{ABM-report}, a process which is mainly sensitive
to the vector NC form factors. A recent determination of  $\mu_s$ has been obtained by 
the HAPPEX experiment at TJLab~\cite{Happex}, also combined with previous 
measurements at BATES~\cite{Sample}. Hence one can hope to get the relevant information
on $g_A^s$ from the future/planned neutrino experiments, which appear to be a unique 
tool for an unambiguous determination of $\Delta s$.

\end{document}